\documentclass[showpacs,nofootinbib,preprintnumbers,prd,superscriptaddress,twocolumn]{revtex4}

\usepackage{amsmath}
\usepackage{amsfonts}
\usepackage{amssymb}
\usepackage{graphicx}
\usepackage[titletoc]{appendix}
\usepackage{color}
\usepackage{hyperref}
\usepackage{cleveref}
\usepackage[rightcaption]{sidecap}
\usepackage{subfigure}
\usepackage{comment}

\usepackage{mathtools}

\usepackage{dcolumn}

\usepackage{array}
\usepackage{ctable}
\usepackage{multirow}
\usepackage{siunitx}
\usepackage{longtable}
\usepackage{tabularx}
\usepackage{booktabs}

\graphicspath{{Graphics/}}

\def\be{\begin{equation}}
\def\ee{\end{equation}}
\def\bea{\begin{eqnarray}}
\def\eea{\end{eqnarray}}

\definecolor{vividviolet}{rgb}{0.62, 0.0, 1.0}
\definecolor{amaranth}{rgb}{0.9, 0.17, 0.31}
\definecolor{palatinateblue}{rgb}{0.15, 0.23, 0.89}
\definecolor{brightpink}{rgb}{1.0, 0.0, 0.5}
\definecolor{cornflowerblue}{rgb}{0.39, 0.58, 0.93}
\definecolor{deepcarminepink}{rgb}{0.94, 0.19, 0.22}
\definecolor{radicalred}{rgb}{1.0, 0.21, 0.37}

\hypersetup{ linktoc=all,
    colorlinks, linkcolor={palatinateblue},
    citecolor={brightpink}, urlcolor={amaranth}
}

\begin{document}

\title{Superhorizon entanglement from inflationary particle production}

\author{Alessio Belfiglio}
\email{alessio.belfiglio@unicam.it}
\affiliation{School of Science and Technology, University of Camerino, Via Madonna delle Carceri, Camerino, 62032, Italy.}
\affiliation{Istituto Nazionale di Fisica Nucleare (INFN), Sezione di Perugia, Perugia, 06123, Italy.}

\author{Orlando Luongo}
\email{orlando.luongo@unicam.it}
\affiliation{School of Science and Technology, University of Camerino, Via Madonna delle Carceri, Camerino, 62032, Italy.}
\affiliation{Istituto Nazionale di Fisica Nucleare (INFN), Sezione di Perugia, Perugia, 06123, Italy.}
\affiliation{SUNY Polytechnic Institute, 13502 Utica, New York, USA.}
\affiliation{INAF - Osservatorio Astronomico di Brera, Milano, Italy.}
\affiliation{Al-Farabi Kazakh National University, Al-Farabi av. 71, 050040 Almaty, Kazakhstan.}

\author{Stefano Mancini}
\email{stefano.mancini@unicam.it}
\affiliation{School of Science and Technology, University of Camerino, Via Madonna delle Carceri, Camerino, 62032, Italy.}
\affiliation{Istituto Nazionale di Fisica Nucleare (INFN), Sezione di Perugia, Perugia, 06123, Italy.}

\begin{abstract}
   We investigate entanglement generation between the sub- and super-Hubble modes of inflaton fluctuations, in the context of particle production from perturbations during inflation. We consider a large-field inflationary scenario where inflation is driven by a vacuum energy symmetry breaking potential and the scalar inflaton field is nonminimally coupled to spacetime curvature. In particular, we focus on the slow-roll phase, adopting a quasi-de Sitter scale factor to properly account for the presence of perturbations and computing the pair production probability associated to the coupling between the inflaton and spacetime inhomogeneities. The interaction Lagrangian at first order is constructed from inhomogeneities induced by the inflaton dynamics, and the initial Bunch-Davies vacuum state of the field evolves under the action of such Lagrangian. In this framework, we quantify the total amount of entanglement via the von Neumann entropy of the reduced density operator for superhorizon modes, tracing out sub-Hubble degrees of freedom. We then compare these outcomes with entanglement production for quadratic chaotic inflation and for a small-field quadratic hilltop scenario, preserving field-curvature coupling in both cases and pointing out the main differences between large and small-field approaches. We show that the amount of entanglement entropy arising from such geometric production grows rapidly in slow-roll regime and that it is typically higher in large-field scenarios. We also discuss our outcomes in light of recent findings for the squeezing entropy of cosmological perturbations and cubic nonlinearities in de Sitter space.
\end{abstract}


\maketitle
\tableofcontents

\section{Introduction}\label{intro}

In recent years, quantum entanglement has been widely recognized as a relevant tool in the context of quantum field theory and gravitation, with important applications spanning from black hole physics \cite{bhol} and anti-de Sitter/conformal field theory  correspondence \cite{ads1,ads2, ads3}, up to de Sitter space dynamics \cite{ds1, ds2, ds3, ds4, ds5}. In particular, de Sitter space plays a key role in modeling the very early universe evolution, since a primordial phase of strong acceleration known as \emph{inflation} \cite{infl1, infl2, infl3} has been proposed to solve the main issues related to the standard Big Bang paradigm \cite{wein,uzan}.

All successful inflationary models should properly deal with the quantum fluctuations of all fields involved, which are thought to be the fundamental seeds for structure formation in our universe \cite{pert1,pert2}. Such fluctuations are typically studied in Fourier space: to leading order, each Fourier mode evolves independently, obeying a harmonic oscillator equation with time-dependent mass. The Hubble horizon then emerges as a natural scale to describe the dynamics of inflaton fluctuations, which typically oscillate in time on sub-Hubble scales, while becoming frozen on super-Hubble ones \cite{rio}.

The corresponding \emph{quantum-to-classical transition} of inflaton fluctuations, i.e., how short wavelength quantum fluctuations are stretched by cosmic expansion and lose their quantum nature, is not yet fully understood, although some headway has been made in the last decades by investigating the effects of squeezing and decoherence \cite{sq1,sq2,sq3,sq4,sq5,sq6,dec1,dec2,dec3,dec4, dec5, meas1, meas2}, also resorting in some cases to open quantum system techniques\footnote{One can, for example, assume that long-wavelength fluctuations decohere in presence of an environment of short-wavelength modes \cite{oqs1,oqs2}. Alternatively, a master equation can be derived by considering local interactions involving a pair of comoving detectors and a given quantum field, usually in its ground state \cite{oqs3,oqs4,oqs5,oqs6,oqs7}.}. Recently, the decoherence mechanism for scalar inflaton fluctuations has been revised by taking into account the role of cubic non-Gaussianities \cite{brand20}. Specifically, due to the nonlinear nature of Einstein equations, it was argued that a certain amount of mode-mixing for cosmological perturbations is always present. This implies a mixing between the sub- and super-Hubble modes, which leads to decoherence of the reduced density matrix of each subsystem. Interestingly, such nonlinearities seem to produce the most relevant contribution to the entropy of cosmological perturbations, surpassing the entropy associated to the squeezed vacuum (see also \cite{bra92,bra93,gasp1,gasp2}). In other words, the entanglement entropy due to cubic interactions dominates over the one due to the (quadratic) squeezing Hamiltonian, provided inflation lasts for a sufficiently long period of time, in agreement with observations \cite{planck}. 

Cubic interactions of density perturbations provide us with a \emph{lower bound} on the amount of entanglement entropy for scalar perturbations. Indeed, such gravitational nonlinearities are inevitably present in most inflationary models. The corresponding entanglement is usually quantified by means of the von Neumann entropy for the reduced density operator corresponding to super-Hubble modes.  However, in recent years it was argued  that inflationary particle production may work as a noise and affect entanglement generation across the Hubble scales \cite{oqs7}. 

Motivated by the above picture, we here investigate the dynamics of superhorizon entanglement associated with geometric particle production \cite{fri,ces} in a single-field inflationary scenario. Geometric production during inflation is typically studied resorting to a perturbative approach\footnote{This mechanism is conceptually different from the so-called gravitational particle production (GPP) from vacuum \cite{gpp1,gpp2,gpp3,gpp4}, associated with cosmological expansion in unperturbed spacetimes and widely studied in different cosmological settings, see e.g. \cite{gpp5,gpp6,gpp7,gpp8,gpp9,gpp10}. However, we will see that geometric production is also affected by GPP, via the Bogoliubov transformations that connect \emph{in} and \emph{out} asymptotic states.}, where the presence of spacetime inhomogeneities is traced back to the quantum fluctuations of the inflaton field and the interacting Lagrangian is constructed by coupling metric perturbations to the energy-momentum tensor of a given quantum field. In particular, the energy-momentum tensor for the inflationary fluctuations of a scalar field has been considered in \cite{bel1}, showing that geometric production may lead to a non-negligible amount of entanglement for particle pairs on super-Hubble scales. Here we extend this approach by studying entanglement entropy generation across the horizon, during the slow-roll regime. We adopt the usual momentum-space entanglement techniques proposed in \cite{msent1,msent2,msent3} and then generalized to de Sitter space\footnote{The here-studied geometric production from inflaton fluctuations can be straightforwardly generalized to the case of a spectator scalar field, which is discussed in Ref. \cite{msent4} assuming a cubic self-interaction term. As argued by the authors, the entropy growth for such spectator field may also mimic entanglement generation for inflationary tensor modes.} in \cite{lel,msent4}, focusing on the pair production probability arising from scalar spacetime perturbations. In so doing, we start from a large-field model of inflation driven by a symmetry breaking potential transporting vacuum energy, also including a nonminimal coupling term between the inflaton field and the scalar spacetime curvature. We compute the von Neumann entropy associated to the reduced density operator for super-Hubble modes and we then compare these outcomes with the widely-studied chaotic quadratic potential \cite{qchaos1,qchaos2,qchaos3} and the small-field quadratic hilltop one \cite{hill1,hill2}, preserving in both cases the nonminimal coupling term with the scalar curvature. For large-field models, we will show that the total amount of entanglement at the end of the slow-roll phase is typically non-negligible. This implies that particle production during the inflationary regime may significantly alter the process of entanglement generation for inflaton fluctuations. Last but not least, we discuss the scale-dependence of entanglement entropy, which represents a peculiar issue of momentum-space calculations. In particular, an ultraviolet (UV) cutoff scale associated with the Planck mass $M_{\rm pl}$ is required, and we also debate limitations on the infrared (IR) scales.

This work is organized as follows. In Sec. \ref{sec2}, we introduce the nonminimally coupled scalar inflaton field and  its quantum fluctuations. In Sec. \ref{sec3}, we review the mechanism of particle production from inhomogeneities. In Sec. \ref{sec4}, we quantify the superhorizon entanglement generated in this process, obtaining the von Neumann entropy associated with super-Hubble modes and comparing these outcomes with a quadratic chaotic model and a small-field quadratic hilltop scenario.  We also discuss our main findings and compare our approach with previous studies. Finally,  Sec. \ref{sec6} is devoted to conclusions and perspectives. We use natural units $c=\hbar=1$ throughout the paper.

%%%%%%%%%%%%%%%%%%%%%%%%%%%%%%%%%%%%%%%%%%%%%%%%%%%%%%%%%%%%%%%%%%%%%%%%%%%%%%%

\section{Inflationary warm up} \label{sec2}

We consider a scalar inflaton field $\phi$, nonminimally coupled to the scalar curvature of spacetime $R$. The corresponding Lagrangian density is written as
\begin{equation}\label{lagrinf}
    \mathcal{L}=\frac{1}{2} \left[ g^{\mu \nu} \phi_{, \mu} \phi_{,\nu}- \xi R \phi^2  \right]- V(\phi),
\end{equation}
where $\xi$ is the field-curvature coupling constant and the potential $V(\phi)$ is left unspecified for the moment. The universe expansion during inflation can be modeled by a spatially flat Friedmann-Robertson-Walker  background, whose line element in cosmic time $t$ reads
$ ds^2=dt^2-a^2(t) d{\bf x}^2$.

As usual, we introduce conformal time $\tau= \int dt/a(t)$ in order to simplify the dynamics of the inflaton field during slow-roll \cite{rio, brand20}. In particular, we can write the unperturbed metric tensor as
$g_{\mu \nu}= a^2(\tau) \eta_{\mu \nu}$, 
where $\eta_{\mu \nu}$ is the Minkowski tensor. 

In conformal time, introducing the \emph{effective potential},
\be \label{effpot}
V^{\rm eff}(\phi, R) \equiv V(\phi)+ \frac{1}{2} \xi R \phi^2,
\ee
the zero-order equation of motion for the inflaton field takes the form
\be \label{zerocomp}
\frac{1}{\sqrt{-g}} \partial_{\mu} \left( \sqrt{-g} g^{\mu \nu} \partial_{\nu} \phi \right)+ V^{\rm eff}_{, \phi}=0,
\ee
with $V^{\rm eff}_{, \phi} \equiv \partial V(\phi)/\partial \phi$ and $g$ is the determinant of the metric tensor.
%%%%%%%%%%%%%%%%%%%%%%%%%%%%%%%%%%%%%%%%%%%%%%%%%%%%%%%%%%%%%%%%%%%%%%%%%%%%%%%%

\subsection{Inflaton quantum fluctuations} \label{sec2B}

We now introduce quantum fluctuations in the above framework. We start by considering the usual ansatz for the inflaton field \cite{rio,bel1}
\be \label{split}
\phi({\bf x},\tau)=\phi_0(\tau)+ \delta \phi ({\bf x},\tau),
\ee
where the background homogeneous contribution $\phi_0$ has been isolated from the perturbing quantum fluctuations $\delta \phi$. The most general metric tensor describing scalar perturbations to linear order can be expressed in the form
\be \label{metrper}
g_{\mu \nu}= a^2(\tau) \begin{pmatrix}
    (1+2\Phi) & \partial_i B \\
    \partial_i B & -\left( (1-2\Psi)\delta_{ij}+ D_{ij} E  \right) \\
\end{pmatrix},
\ee
where $\Phi$, $\Psi$, $B$ and $E$ are scalar quantities and $D_{ij}\equiv \partial_i \partial_j - \frac{1}{3} \delta_{ij} \nabla^2$. 

Hereafter, we adopt the \emph{longitudinal}, or conformal Newtonian gauge\footnote{Another common choice is the so-called \emph{comoving gauge}, which is employed in \cite{brand20} to study momentum-space entanglement. See, for example, \cite{infl3} for an introduction to the most popular gauge choices associated to cosmological perturbations.}, equivalent to set $E=B=0$. Moreover, when the fluctuation matter source has no anisotropic stress, as for the scalar inflaton, we can further set $\Phi=\Psi$. In Fourier space, we write perturbation modes as \cite{pert1}
\be \label{pertmod}
\Psi({\bf x}, \tau)= \Psi_k(\tau) e^{i {\bf k}\cdot {\bf x}},
\ee
and, similarly, we can expand quantum fluctuations as
\be \label{qfesp}
\hat{\delta \phi} ({\bf x}, \tau)= \frac{1}{(2 \pi)^{3/2}} \int d^3k \left( \hat{a}_k \delta\phi_k e^{i {\bf k}\cdot {\bf x}} + \hat{a}_k^\dagger \delta\phi_k^* e^{-i {\bf k}\cdot {\bf x}} \right), 
\ee
satisfying the usual canonical commutation relations, $
[\hat{a}_k, \hat{a}_{k^\prime}^\dagger]= \delta^{(3)} ({\bf k}-{\bf k^\prime})$.

Following Refs. \cite{rio,bel1} and dropping the subscript $0$ for clearness, the first-order perturbed equations become
\begin{align} \label{perteq}
&\delta \phi_k^{\prime \prime}+ 2 \mathcal{H} \delta \phi_k^\prime+ k^2 \delta \phi_k-4 \Psi_k^\prime \phi^\prime \notag \\
&=- \xi \left( -2k^2 \Psi_k-6\Psi_k^{\prime \prime}-24 \mathcal{H} \Psi_k^\prime-12 \frac{a^{\prime \prime}}{a} \Psi_k  \right) \phi \notag \\
&\ \ \ - \left( V^{\rm eff}_{, \phi \phi} \delta \phi_k + 2 \Psi_k V^{\rm eff}_{, \phi}  \right) a^2,
\end{align}
where the prime denotes derivatives with respect to conformal time and $\mathcal{H} \equiv a^\prime/a$. If $\lvert \xi \rvert \ll 1$, we can exploit the slow-roll condition and the $(0,i)$ component of perturbed Einstein's equations
\be \label{perturb}
\Psi_k^\prime+\mathcal{H} \Psi_k= \epsilon \mathcal{H}^2 \frac{\delta \phi_k}{\phi^\prime},
\ee
where $\epsilon$ is a slow-roll parameter specified later on, to simplify Eq. \eqref{perteq} as
\be \label{perteqs}
\delta \phi_k^{\prime \prime}+2 \mathcal{H} \delta \phi_k^\prime+\left[ k^2 + V^{\rm eff}_{, \phi \phi}a^2 +6 \epsilon \mathcal{H}^2 \right] \delta \phi_k = 0.
\ee

%%%%%%%%%%%%%%%%%%%%%%%%%%%%%%%%%%%%%%%%%%%%%%%%%%%%%%%%%%%%%%%%%%%%%%%%%%%%%%%%

\subsection{Inflaton modes in a quasi-de Sitter background} \label{sec2C}

As mentioned in Sec. \ref{intro}, inflation leads to a de Sitter phase during which the universe experiences a period of intense acceleration. However, the inflationary epoch cannot precisely manifest an \emph{exact} de Sitter phase, since small deviations are inevitably present as consequence of the inflaton dynamics. To accomplish these slight deviations, we here adopt a \emph{quasi-de Sitter} background evolution \cite{rio}
\be \label{quasids}
a(\tau)= -\frac{1}{H_I \tau^{(1+\epsilon)}},\ \ \ \ \epsilon \ll 1,
\ee
where $\tau <0$ and $H_I$ is the Hubble parameter during inflation, while $\epsilon$ can be identified as a small and constant \emph{slow-roll parameter}. 

Inserting Eq. \eqref{quasids} into Eq. \eqref{perteqs} and rescaling the field by 
$\delta \chi_k= \delta \phi_k a$, we obtain
\be \label{modes}
\delta \chi_k^{\prime \prime}+ \left[ k^2-\frac{1}{\tau^2} \left( \left( 1-6\xi\right)(2+3\epsilon)+6\epsilon- \frac{ V^{\rm eff}_{\phi \phi}}{H_I^2} \right)    \right] \delta \chi_k=0,
\ee
where we also computed the scalar curvature in conformal time, $R=6a^{\prime \prime}/a^3$, and noted that $a^{\prime \prime}/a \simeq (2+3\epsilon)/\tau^2$. Eq. \eqref{modes} admits solutions 
\be \label{hanksol}
\delta \chi_k(\tau)= \sqrt{-\tau} \left[ c_1(k) H_{\nu}^{(1)}(-k\tau)+c_2(k) H_{\nu}^{(2)}(-k\tau)  \right],
\ee
where $H_{\nu}^{(1)}$ and $H_{\nu}^{(2)}$ are Hankel functions and
\be \label{hankind}
\nu= \sqrt{\frac{1}{4}+(1-6\xi)(2+3\epsilon)+6\epsilon-V^{\rm eff}_{,\phi \phi}/H_I^2}.
\ee
The integration constants $c_1(k)$ and $c_2(k)$ can be determined by choosing the state of the field when inflation starts. A common ansatz consists in employing the \emph{Bunch-Davies vacuum state} \cite{bunch1,bunch2,bunch3}, that corresponds to impose the boundary condition
\be \label{asymp}
\delta \chi_k  \xrightarrow[\tau \rightarrow -\infty]{} \frac{e^{-ik\tau}}{\sqrt{2k}}.
\ee
This choice implies  $c_1(k)=\sqrt{\pi}e^{i(\nu +\frac{1}{2})\frac{\pi}{2}}/2$ and $c_2(k)=0$, so the original fluctuation modes take the form
\be \label{influct}
\delta \phi_k (\tau)= \frac{\sqrt{-\pi\tau}}{2} e^{i\left( \nu+ \frac{1}{2}\right) \frac{\pi}{2}}  H^{\left(1\right)}_{\nu}\left(-k\tau\right) /a(\tau).
\ee
Our primary focus will be to compute entanglement between sub- and super-Hubble modes. Thus, we exploit the asymptotic behavior of Hankel functions
\begin{align}
&H^{\left(1\right)}_{\nu}\left(x\gg1\right) \simeq\sqrt{\frac{2}{\pi x}}e^{i\left(x-\frac{\pi}{2}\nu-\frac{\pi}{4}\right)},\\
&H^{\left(1\right)}_{\nu}\left(x\ll1\right)\simeq \sqrt{\frac{2}{\pi}}e^{-i\frac{\pi}{2}}2^{\left(\nu-\frac{3}{2}\right)}\frac{\Gamma\left(\nu\right)}{\Gamma\left(\frac{3}{2}\right)}x^{-\nu}
\end{align}
to derive the expressions for fluctuations inside and outside the comoving Hubble radius $r_H(\tau)=1/(a(\tau)H_I)$. We then introduce
\begin{align}
&{\emph{ Sub-Hubble\,\,scales}:}\nonumber\\
    &\delta \phi_k^{\rm sub} \simeq \frac{1}{\sqrt{2k}}  \frac{e^{i \left( \nu+ \frac{1}{2} \right) \frac{\pi}{2}}\ e^{i\left( -k\tau-\frac{\pi}{2}\nu-\frac{\pi}{4}  \right)}}{a(\tau)}, \label{subfluc}\\
    \,\nonumber\\
&{\emph{ Super-Hubble\,\,scales}:}\nonumber\\
    &\delta \phi_k^{\rm super} \simeq e^{i\left(\nu-\frac{1}{2}\right)\frac{\pi}{2}}2^{\left(\nu-\frac{3}{2}\right)}\frac{\Gamma\left(\nu\right)}{\Gamma\left(\frac{3}{2}\right)}\frac{H_{I}}{\sqrt{2k^{3}}}\left(\frac{k}{aH_{I}}\right)^{\frac{3}{2}-\nu}. \label{supfluc}
\end{align}
Remarkably, we notice that fluctuations oscillate on sub-Hubble scales, $k \gg aH_I$, while being nearly frozen after crossing the horizon. Later on, we will specify these calculations to some relevant inflationary potentials, in order to derive the dynamics of the corresponding inflaton fluctuations and quantify entanglement generation during the slow-roll phase. First, we briefly recall the mechanism of particle production from inhomogeneities during inflation.

%%%%%%%%%%%%%%%%%%%%%%%%%%%%%%%%%%%%%%%%%%%%%%%%%%%%%%%%%%%%%%%%%%%%%%%%%%%%%%%%

\section{Geometric particle production} \label{sec3}

Particle production from spacetime inhomogeneities has been proposed some time ago \cite{fri, ces} as an alternative mechanism to the widely-studied GPP scenario. More precisely, 

\begin{itemize}
    \item[-] the presence of space-dependent perturbations in an expanding background is expected to enhance the total amount of particles produced,
    \item[-] such perturbations break the space translation symmetry \cite{bel2} and, accordingly, ``geometric" production is not limited to pair production with opposite momenta.
\end{itemize}

\noindent Assuming a perturbed background, with
\be \label{pertmet}
g_{\mu \nu}= a^2(\tau) \left( \eta_{\mu \nu}+ h_{\mu \nu} \right),\ \ \ \ \lvert h_{\mu \nu} \rvert \ll 1,
\ee
we can write the first-order interaction Lagrangian density describing the coupling between perturbations and a given quantum field in the form
\be \label{intlag}
    \mathcal{L}_{I}=-\frac{1}{2}\sqrt{-g_{(0)}}H^{\mu\nu}T^{\left(0\right)}_{\mu\nu},
\ee
where $T_{\mu \nu}^{(0)}$ is the zero-order energy-momentum tensor for the field, $g_{(0)}$ the determinant of the background unperturbed metric tensor and $H_{\mu \nu}= a^2(\tau) h_{\mu \nu}$. In the longitudinal gauge, scalar perturbations associated with $\phi$ take the simple form
\be \label{pertens}
h_{\mu \nu}=\begin{pmatrix}
2 \Psi & 0 & 0 & 0 \\
0 & 2 \Psi & 0 & 0 \\
0 & 0 & 2\Psi & 0 \\
0 & 0 & 0 & 2 \Psi\\[4.5pt]
\end{pmatrix}.
\ee
The perturbation potential $\Psi$ is derived from Eq. \eqref{perturb}, which can be recast in the form 
\be \label{pertslow}
\Psi_k^{\prime \prime}+2\left(\mathcal{H}-\frac{\phi^{\prime \prime}}{\phi^\prime} \right) \Psi_k^\prime+2\left( \mathcal{H}^\prime-\mathcal{H} \frac{\phi^{\prime \prime}}{\phi^\prime}\right) +k^2 \Psi_k=0,
\ee
where we also exploited the mode decomposition of Eq. \eqref{pertmod}. Moreover, since we focus on the dynamics of inflaton fluctuations\footnote{An alternative approach involves the introduction of \emph{spectator fields}. See, for example, Refs. \cite{msent4, boya} for some studies on entanglement generation associated to spectator fields during inflation. Such fields may mimic the dynamics of tensor modes in inflation, as argued in \cite{msent4}.}, we assume
\begin{align} \label{zerotens}
T_{\mu \nu}^{(0)}=& \partial_{\mu} \delta \phi\  \partial_{\nu}\delta \phi-\frac{1}{2} g_{\mu \nu}^{(0)} \left[ g^{\rho \sigma}_{(0)}\  \partial_{\rho} \delta \phi\  \partial_{\sigma} \delta \phi - V(\delta \phi)  \right] \notag \\ 
&- \xi \left[ \nabla_{\mu} \partial_{\nu}- g_{\mu \nu}^{(0)} \nabla^\rho \nabla_\rho+R_{\mu \nu}^{(0)}-\frac{1}{2} R^{(0)} g_{\mu \nu}^{(0)}   \right] (\delta \phi)^2.
\end{align}
Introducing now the interaction Hamiltonian density as $\mathcal{H}_I$, it can be shown that $\mathcal{L}_I=-\mathcal{H}_I$  holds for a Lagrangian density of the form \eqref{intlag}. This implies that we can write the $S$ matrix at first-order in Dyson's expansion\footnote{A proper definition of the $S$ matrix in curved spacetime is not always straightforward, since it requires the existence of asymptotically flat regions (or, at least, asymptotic adiabatic regimes \cite{gpp2}) to properly define vacuum and particle states. See, for example, Ref. \cite{bel1} for a discussion on these issues in the context of inflationary particle production from inhomogeneities. At the same time, the fast growth of the inflationary scale factor may break the perturbative approach at some point. We will come back to this further issue in Sec. \ref{sec5}.} as
\be \label{smat}
\hat{S} \simeq 1 + i \hat{T} \int d^4x \mathcal{L}_I.
\ee
We will focus on particle pair production for inflationary potentials of the form $V \propto \phi^{2n}$ ($n=1,2$). The corresponding probability amplitude for pair creation reads
\begin{align} \label{compact}
\mathcal{C}_{{\bf p}_1,{\bf p}_2} & \equiv \langle {\bf p}_1,{\bf p}_2 \lvert \hat S \rvert 0 \rangle \notag \\ 
& = -\frac{i}{2} \int d^4x\  2a^2\big(A_0({\bf x}, \tau)
+A_1({\bf x},\tau) \notag \\
&\ \ \ \ \ \ \ \ \ \ \ \ \ \ \ \ \ \ \ \ \ \ \ \ \ \ +A_2({\bf x},\tau)+A_3({\bf x},\tau) \big),
\end{align}
where
\begin{align} \label{Atime}
A_0({\bf x},\tau)= 2 \Psi \bigg[&\partial_0 \delta \phi_{p_1}^*\  \partial_0 \delta \phi_{p_2}^* \notag \\
&-\frac{1}{2} \big( \eta^{\rho \sigma} \partial_\rho \delta \phi_{p_1}^* \ \partial_\sigma \delta \phi_{p_2}^*-a^2 V(\delta \phi) \big) \notag \\
    &- \xi \bigg(\partial_0 \partial_0-\frac{a^\prime}{a}\partial_0 -\eta^{\rho \sigma}\partial_\rho \partial_\sigma \notag \\
    &\ \ \ \ \ \ \ -3\left( \frac{a^\prime}{a} \bigg)^2   \right) \delta \phi_{p_1}^* \delta\phi_{p_2}^* \bigg] e^{-i({\bf p_1}+{\bf p_2})\cdot {\bf x}}\notag\\
\end{align}
and
\begin{align} \label{Aspace}
    A_i({\bf x},\tau)=2 \Psi \bigg[&\partial_i \delta \phi_{p_1}^*\  \partial_i \delta \phi_{p_2}^*\notag \\
    & +\frac{1}{2} \big( \eta^{\rho \sigma} \partial_\rho \delta \phi_{p_1}^* \ \partial_\sigma \delta \phi_{p_2}^*-a^2V(\delta \phi) \big)\notag \\
    &- \xi \bigg(\partial_i \partial_i+\frac{3a^\prime}{a}\partial_0+ \frac{2a^{\prime \prime}}{a}+\eta^{\rho \sigma}\partial_\rho \partial_\sigma \notag \\
    &\ \ \ \ \ \ \ \ -\left( \frac{a^\prime}{a} \right)^2  \bigg) \delta \phi_p^* \delta\phi_q^* \bigg] e^{-i({\bf p_1}+{\bf p_2})\cdot {\bf x}},\notag\\
\end{align}
for $i=1,2,3$. For each particle pair, the final state can be written in the form
\be \label{finstat}
\lvert \Psi \rangle = \hat{S} \lvert 0_{{\bf p}_1}; 0_{{\bf p}_2} \rangle = \mathcal{N} \left( \lvert 0_{{\bf p}_1}; 0_{{\bf p}_2} \rangle + \frac{1}{2} \mathcal{C}_{{\bf p}_1,{\bf p}_2} \lvert 1_{{\bf p}_1};1_{{\bf p}_2} \rangle  \right),
\ee
where the normalization factor $\mathcal{N}$ is derived as usual by imposing $\langle \Psi \rvert \Psi \rangle=1$. The total number density of particles arising from the interacting Lagrangian, Eq. \eqref{intlag}, at second perturbative order is then
\begin{align} \label{numbdens}
N^{(2)}(\tau)= \frac{a^{-3}(\tau)}{\left(2\pi  \right)^{3}} \int d^3p_1\  d^3p_2\  & \lvert \mathcal{C}_{{\bf p}_1,{\bf p}_2}  \rvert^2 \notag \\ &\times \left( 1+ \lvert \beta_{p_1} \rvert^2+ \lvert \beta_{p_2} \rvert^2 \right).
\end{align}
In Eq. \eqref{numbdens}, we  introduced the Bogoliubov coefficients $\beta_{p_1}$ and $\beta_{p_2}$, which can be derived from the dynamics of the field modes $\delta \phi_k$ in the asymptotic regions $\tau \rightarrow \pm \infty$. It can be shown that Bogoliubov coefficients are zero in a purely de Sitter spacetime \cite{gpp2}, while in the context of GPP they can be computed assuming a rapid transition from inflation to the radiation dominated era \cite{gpp10,boya}. In these models, the quasi de Sitter scale factor of Eq. \eqref{quasids} should be modified in order to guarantee a continuous transition between the two epochs, thus avoiding divergences. The process of gravitational production is more effective during the nonadiabatic inflationary expansion of the universe, while adiabaticity is gradually recovered before matter radiation equality, so that a proper definition of particle states is again possible at sufficiently large $\tau$.  Bogoliubov coefficients can also describe the squeezing of cosmological perturbations during inflation \cite{brand20, bra92, bra93, gasp1, gasp2}, even if this approach has been recently criticized due to possible ambiguities that may arise in defining squeezed states for quantum fields in expanding backgrounds \cite{sq6}. We will return to this issue in Sec. \ref{sec5}, while in the following we are going to compute superhorizon entanglement associated to geometric particle production during the slow-roll phase of inflation. From now on, we will neglect the contribution of Bogoliubov coefficients to particle creation from inhomogeneities, thus focusing on the pure geometric contribution described by the probability amplitude of Eq. \eqref{compact}.

%%%%%%%%%%%%%%%%%%%%%%%%%%%%%%%%%%%%%%%%%%%%%%%%%%%%%%%%%%%%%%%%%%%%%%%%%%%%%%%

\section{Superhorizon entanglement entropy} \label{sec4}

The degrees of freedom of any interacting quantum field theory are entangled in momentum space \cite{msent1}. In particular, inflationary perturbations are mostly studied in momentum space, as the properties of momentum modes are typically probed in experimental contexts \cite{planck}.  

Specifically, during inflation the (comoving) Hubble radius $r_H$ naturally divides the total Hilbert space of states into sub-Hubble  and  super-Hubble subspaces, respectively \cite{brand20}. This implies that the entanglement entropy of inflaton fluctuations can be identified as the von Neumann entropy associated with the reduced density matrix of one of these two subsystems. 

Given the generic probability amplitude for particle production up to the first order in $h$,
\be \label{matel}
\mathcal{C}_{n,N} = \langle n, N \lvert \left( -i \int_{\tau_0}^\tau d\tau^\prime \mathcal{H}_I (\tau^\prime) \right) \rvert 0,0 \rangle + \mathcal{O}(h^2),
\ee
where $n$ is the number of particles in the subsystem traced out and $N$ the particles in the main system, we can write the von Neumann entropy of the reduced state as
\be \label{entposisp}
\mathcal{S}_{\rm ent}= - \sum_{n, N \neq 0} \lvert \mathcal{C}_{n,N} \rvert^2 \left(\ln\  \lvert \mathcal{C}_{n,N} \rvert^2 -1 \right)+ \mathcal{O}(h^3).
\ee
This result can be re-expressed in momentum space by the substitution
$\sum_{n,N \neq 0} \rightarrow \sum_{\{{\bf p}_i \} \gtrless \mu}$, where $\mu$ stands for a generic momentum cutoff scale, obtaining the following entropy per unitary volume
\begin{align} \label{entmom}
\mathcal{S}_{\rm ent}=&- \int_{\{{\bf p}_i \} \gtrless \mu} \prod_{i}^d d^3p_i \left[ \lvert \mathcal{C}_{\{{\bf p}_i \} \gtrless \mu} \rvert^2 \left( \ln\  \lvert \mathcal{C}_{\{{\bf p}_i \} \gtrless \mu} \rvert^2 -1\right)  \right] \notag \\ &+ \mathcal{O}(h^3).
\end{align}
In case of entanglement associated with inflationary fluctuations, the Hubble radius is the natural separation scale for field modes. Hence, we can set $\mu \equiv r_H(\tau)$ in order to compute entanglement across the Hubble horizon. Moreover, one usually assumes the space of super-Hubble modes to be the system under analysis, while the sub-Hubble space is treated as the ``bath" we integrate over. 

Focusing now on perturbative particle production from inhomogeneities, we are interested in amplitudes of the form
\be \label{matcou}
\mathcal{C}_{\{{\bf p}_i\}}= \langle {\bf p}_1 {\bf p}_2 \lvert \left( i \int_{\tau_0}^\tau d\tau^\prime \mathcal{L}_I (\tau^\prime) \right) \rvert 0, 0 \rangle + \mathcal{O}(h^2),
\ee
where $\mathcal{L}_I$ is given by Eq. \eqref{intlag} and we are focusing on quadratic terms only.

In our treatment, the total amount of superhorizon entanglement is thus obtained by picking one mode on super-Hubble scales and the other on sub-Hubble ones, namely
\be \label{setmom}
\{ {\bf p}_i \}= \begin{cases}  a(\tau_i)H_I < \lvert {\bf p}_1 \rvert < a(\tau) H_I, \\ a(\tau) H_I < \lvert {\bf p}_2 \rvert < a(\tau) M_{\rm pl}.   \end{cases}
\ee
In Eq. \eqref{setmom}, we introduced the usual UV cutoff for comoving momenta in terms of the Planck mass, $M_{\rm pl}$. At the same time, our IR cutoff consists in neglecting all modes whose wavelength exceeds the Hubble radius at the beginning of inflation. Consequently, with this choice all super-Hubble modes are created during the quasi-de Sitter expansion, thus allowing a causal generation mechanism for cosmological perturbations \cite{brand20}.

In the following, we specify the here-presented approach to some relevant inflationary potentials.

\subsection{Choosing the inflationary potentials}

\begin{figure}
    \centering
    \includegraphics[scale=0.8]{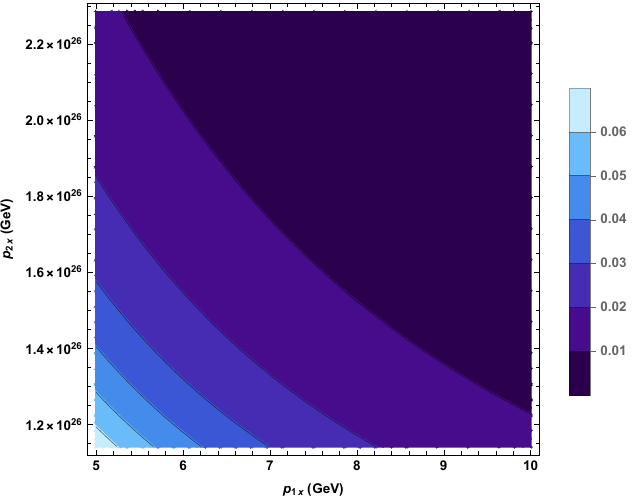}
    \caption{Superhorizon entanglement entropy $\mathcal{S}_{\rm ent}$ at time $\tau_f$ from geometric particle production, assuming a quartic self-coupling inflationary potential. The entropy is plotted as function of the super-Hubble mode $p_{1x}$ and the sub-Hubble mode $p_{2x}$. The other momentum components are set to zero, for simplicity. We assume $\phi(\tau_i)=5$ $M_{\rm pl}$, $\lambda=10^{-15}$, $\epsilon=10^{-3}$, $\xi=10^{-4}$ and $\tau^*=\tau_0/1000=-10^{-3}$ GeV$^{-1}$.}
    \label{figquart}
\end{figure}
%%%%%%%%%%%%%%%%%%%%%%%%%%%%%%%%%%%%%%%%%%%%%%%%%%%%%%%%%%%%%%%%%%%%%%%%%%%%%%%%

Recent results from the Planck satellite have imposed stringent constraints on the potentials that can be feasibly adopted to drive the inflationary epoch \cite{planck}. Power-law potentials, in particular, seem to be disfavored by observations, albeit the curvature coupling seems to resort their use in inflationary stages. Analogously, chaotic and hilltop potentials still appear to be plausible options to describe inflation. Motivated by these results, in order to quantify superhorizon entanglement arising from geometric particle production, we focus on three specific potentials, as discussed below.

\begin{itemize}
    \item[-] We start from a quartic self-coupling interaction, selecting the potential
\be \label{quarticpot}
V(\phi)= \frac{\lambda}{4} \left( \phi^2- v^2 \right)^2,
\ee
where $v$ is the vacuum expectation value of the inflaton field. Inflationary predictions associated to a nonminimal quartic potential have been studied in detail, see e.g. \cite{qchaos2,qchaos3,tsuqua}. Moreover, this potential has been recently proposed in the context of vacuum energy cancellation during inflation \cite{carl}.

In a large-field scenario, we can safely set $V(\phi) \simeq \lambda \phi^4/4$ during slow-roll. Accordingly, the slow-roll background equation for the inflaton field can be expressed in conformal time as
\be \label{backquar}
3 \mathcal{H} \phi^\prime= - \left(\lambda \phi^3+ \xi R \phi  \right)a^2,
\ee
that, for a quasi-de Sitter evolution, gives
\be \label{backq}
\phi^\prime- \frac{3\xi(2+3\epsilon)}{(1+\epsilon)\tau}\phi= \frac{\lambda}{2(1+\epsilon)H_I^2 \tau}\phi^3.
\ee
Eq. \eqref{backq} is solved resorting to the usual techniques for \emph{Bernoulli differential equations}.

Once derived the background dynamics, we can obtain the fluctuation modes for the inflaton field. For simplicity, during slow-roll we make the substitution
\be \label{avgback}
\phi^2 \rightarrow \frac{\int_{\tau_i}^{\tau_f} \phi^2 d\tau}{ \tau_f-\tau_i} \equiv \langle \phi^2 \rangle,
\ee
thus identifying $\phi^2$ with its mean value. The final time $\tau_f$ is usually determined by imposing a number $N_{\rm tot}$ of e-foldings which is sufficient to solve the horizon and flatness problems, e.g. 
\be \label{efold}
N_{\rm tot}= \int dt H(t) \simeq - \int_{\tau_i}^{\tau_f} d\tau/\tau =60.
\ee
Exploiting now Eqs. \eqref{backq}-\eqref{avgback}, the solution for inflaton fluctuations is in the form \eqref{influct}, with
\be \label{indquart}
\ \ \ \ \ \ \ \ \nu \simeq  \sqrt{\frac{1}{4}+(1-6\xi)(2+3\epsilon)+6\epsilon- 3 \lambda\langle \phi^2 \rangle/H_I^2}.
\ee
When dealing with quartic terms, we are interested in probability amplitudes of the form
\begin{align} \label{probamp}
\mathcal{C}_{{\bf p}_1,{\bf p}_2} & \propto \langle {\bf p}_1,{\bf p}_2 \lvert \hat{T} \left[ (\delta \phi)^4 \right] \rvert 0 \rangle \notag \\
& = 12\  \delta \phi_{p_1}^*({\bf x}, \tau) \delta \phi_{p_2}^*({\bf x}, \tau) G_F \langle 0 \rvert 0 \rangle,
\end{align}
where $\hat{T}$ is the time-ordering operator and $\lvert 0 \rangle$ denotes the Bunch-Davies initial vacuum state\footnote{Here we are working with un-normalized states, following the conventions of \cite{gpp2}.}. Such amplitudes clearly diverge, due to the presence of the Feynman propagator $G_F$ and, so, the need of renormalizing self-interacting scalar field theories in curved spacetime is crucial. 

Hence, cancelling poles in the amplitudes requires  appropriate rescaling of the coupling constants. For example, by setting
\begin{subequations}
    \begin{align}
    & \xi_R= \xi-\delta \xi, \label{rn1} \\
    & \lambda_R= \lambda-\delta \lambda, \label{rn2}
    \end{align}
\end{subequations}
where $\xi_R$ and $\lambda_R$ are renormalized constants, we obtain
\begin{align} \label{renor}
\ \ \ \ \ \langle {\bf p}_1,{\bf p}_2 &\lvert \hat{T}\left[(\delta \phi)^4 \right] \rvert 0 \rangle_{\rm ren} \propto\  \delta \phi_{p_1}^*({\bf x}, \tau) \delta \phi_{p_2}^*({\bf x}, \tau) \notag \\
&\  \ \ \ \ \ \ \ \times \left \{ G_F^{\rm fin}+ \frac{i}{8 \pi^2} \left( \xi_R -\frac{1}{6}\right) R \ln \mu \right \} . 
\end{align}
In Eq. \eqref{renor}, the propagator $G_F^{\rm fin}$ is finite in four dimensions \cite{gpp2}, while $\mu$ is an arbitrary mass scale. Rescaling of $\mu$ simply readjusts the relations between bare and renormalized constants, thus implying that measurements are required to fix the $\lambda_R$ and $\xi_R$ magnitudes.\\
%%%%%%%%%%%%%%%%%%%%%%%%%%%%%%%%%%%%%%%%%%%%%%%%%%%%%%%%%%%%%%%%%%%%%%%%%%%%%%%%
\begin{figure}
    \centering
    \includegraphics[scale=0.8]{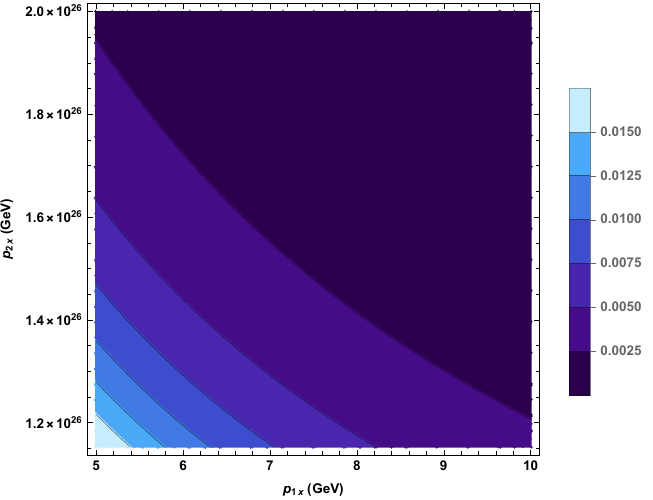}
    \caption{Superhorizon entanglement entropy $\mathcal{S}_{\rm ent}$ at time $\tau_f$ from geometric particle production, assuming a quadratic chaotic potential. The entropy is plotted as function of the super-Hubble mode $p_{1x}$ and the sub-Hubble mode $p_{2x}$.  The other momentum components are set to zero, for simplicity. We assume $\phi(\tau_i)=5$ $M_{\rm pl}$, $\epsilon=10^{-3}$, $\xi=10^{-4}$ and $\tau^*=\tau_0/1000=-10^{-3}$ GeV$^{-1}$. Moreover, the mass of the field is $m=1.34\times 10^{12}$ GeV, in order to obtain the same  expansion law of the quartic case.}
    \label{figquad}
\end{figure}
%%%%%%%%%%%%%%%%%%%%%%%%%%%%%%%%%%%%%%%%%%%%%%%%%%%%%%%%%%%%%%%%%%%%%%%%%%%%%%%%

In case of inflationary particle production, we expect that the amount of particles arising from quartic self-coupling is approximately equal to the total number of pairs obtained from curvature, since the energy scales of the two terms are comparable during slow-roll. Accordingly, from now on we will focus on the field-curvature coupling term, because in this latter case the corresponding probability amplitudes are computed without the need of renormalization techniques.

Even neglecting the quartic contribution in $V(\delta \phi)$, the probability amplitude $\mathcal{C}_{{\bf p}_1,{\bf p}_2}$ of Eq. \eqref{compact} is not straightforward to compute, due to the presence of derivative terms in our interaction Lagrangian. As a simplifying assumption, we first average in time the perturbation potential $\Psi$  during the slow-roll phase, assuming
\be \label{avgpert}
\Psi_k \rightarrow \frac{\int_{\tau_i}^{\tau_f} \psi_k d\tau}{ \tau_f-\tau_i} \equiv \langle \Psi_k \rangle,
\ee
which now behaves as a momentum-dependent coupling term in the interaction Lagrangian. Moreover, we select a time $\tau^* > \tau_i$ and focus on particle production in the interval $[\tau^*, \tau_f]$: by virtue of this choice, we can exploit the simplified solution of Eq. \eqref{supfluc} for super-Hubble modes in the range  $a(\tau_i) H_I < k < a(\tau^*)H_I$. This last assumption inevitably leads to underestimate the total entanglement generated in the slow-roll phase, since we are neglecting all modes that cross the horizon after $\tau^*$. However, these contributions are typically small, as we now discuss.

In Fig. \ref{figquart} we show the superhorizon entanglement entropy arising from geometric particle production at the end of the slow-roll phase, as function of sub- and super-Hubble momentum modes. We specify to the case of particle production along the $x$ direction\footnote{The result is independent of this choice, due to the spherical symmetry of the perturbation.} and observe that a significant amount of entanglement can be produced in such processes. We also notice that the entropy is larger at small momenta: in case of super-Hubble modes, this implies that entanglement increases when approaching the IR cutoff and reflects the bosonic nature of the field under investigation. The same result is obtained when studying entanglement due to cubic nonlinearities \cite{brand20} or spectator scalar fields during inflation \cite{msent4}. Finally, from Eqs. \eqref{Atime}-\eqref{Aspace} we notice that the dominant term in the probability amplitude \eqref{compact} is proportional to the scalar curvature and, assuming particle production across the horizon, it satisfies $\mathcal{C}_{{\bf p}_1,{\bf p}_2}(R) \propto a^3$. Accordingly, entanglement production due to inhomogeneities typically exhibits a fast growth during slow-roll, as consequence of the rapid expansion of the universe. 

\item[-] We now move to the widely-studied chaotic potential
\be 
V(\phi)= \frac{1}{2}m^2 \phi^2,
\ee
where $m$ is the mass of the inflaton field. For this model, entanglement production due to inhomogeneities has been recently investigated in \cite{bel1}, focusing on super-Hubble degrees of freedom. 
In this case, Eq. \eqref{modes} for the field flucuations can be solved exactly, with
\be \label{indquad}
\ \ \ \ \ \ \nu =  \sqrt{\frac{1}{4}+(1-6\xi)(2+3\epsilon)+6\epsilon- m^2/H_I^2}.
\ee
In Fig. \ref{figquad}, we plot the corresponding superhorizon entanglement, i.e., the von Neumann entropy for the reduced state of super-Hubble modes, as function of the momenta $p_{1x}$ (super-Hubble) and $p_{2x}$ (sub-Hubble). We notice that the amount of entanglement produced in this model is smaller but still comparable to the previously studied self-coupling case. This is expected, since the energy of the inflaton field is similar in the two scenarios, due to the choice of initial conditions for the background field $\phi$. For a quadratic chaotic model with nonminimal coupling, the dominant terms for particle production are associated to the mass and the scalar curvature: both gives again $\mathcal{C}_{{\bf p}_1,{\bf p}_2} \propto a^3$, which certifies the fast growth of entanglement entropy during slow-roll.  Moreover, the entropy is still dominated by momenta close to the IR cutoff, thus confirming that in case of scalar fields the entanglement is higher for low-momentum modes.  \item[-]   Finally, we investigate a small-field scenario by selecting the quadratic hilltop potential \cite{hill1}
\be \label{hillpot}
V(\phi)= \Lambda^4\left(1-\frac{\phi^2}{\mu_2^2} \right),
\ee
where $\Lambda^4$ represents the vacuum energy density during inflation and the parameter $\mu_2$ is experimentally constrained by 
\be \label{muconstr}
0.3 < \log_{10}\left( \mu_2/M_{\rm pl} \right) < 4.85
\ee
in the minimally coupled scenario \cite{planck}. Geometric particle production in a nonminimally coupled hilltop model was studied in \cite{hill2}, suggesting that this process may help in alleviating the cosmological constant problem \cite{cc1,cc2}. The corresponding geometric particles represent possible dark matter candidates, under certain conditions. For a quadratic hilltop potential, inflaton fluctuations admit again solutions in the form of Eq. \eqref{influct}, with 
\be \label{indhill}
\ \ \ \ \ \ \ \nu=  \sqrt{\frac{1}{4}+(1-6\xi)(2+3\epsilon)+6\epsilon+ 2\Lambda^4/(\mu_2^2 H_I^2)}.
\ee
However, in this case the energy associated to the inflaton field is typically small during the slow-roll phase, thus implying that entanglement generation due to inflaton fluctuations is necessarily less efficient with respect to large-field approaches. In fact, setting for example $\phi(\tau_i)=1$ GeV and $\mu_2=1$ $M_{\rm pl}$, one finds that the corresponding superhorizon entanglement is around sixty orders of magnitude smaller than the previously studied quartic and quadratic models. This suggests that in small-field models we expect no significant amount of entanglement to be preserved after horizon crossing.
\end{itemize}

%%%%%%%%%%%%%%%%%%%%%%%%%%%%%%%%%%%%%%%%%%%%%%%%%%%%%%%%%%%%%%%%%%%%%%%%%%%%%%%%

\subsection{Theoretical consequences} \label{sec5}

As we have seen, the calculation of entanglement entropy for cosmological perturbations is most easily performed in momentum space. This helps to avoid some technical problems that arise when dealing with position-space entanglement, especially for interacting fields \cite{msent4}.  Moreover, momentum-space entanglement may carry information corresponding to measurable observables in the Cosmic Microwave Background, since the properties of momentum modes are those generally probed by experiments \cite{planck}. 

We have shown that the process of geometric particle production during inflation leads to a non-negligible amount of entanglement entropy in large-field models. Moreover, such entropy rapidly grows during the slow-roll phase and, more generally, for a fast-expanding background. This implies that \emph{perturbative particle production mechanisms can significantly affect the quantum properties of inflationary perturbations}. At the same time, the universe expansion during inflation is expected to squeeze cosmological perturbations after horizon crossing. It was recently shown in Ref. \cite{brand20} that the squeezing entropy of perturbations can be derived by considering cubic interaction terms for fluctuations, that are responsible for reducing the pure density matrix to a mixed one, by suppressing off-diagonal terms. Looking at Eq. \eqref{numbdens}, we notice that the nonperturbative expansion of the background, quantified by the Bogoliubov coefficients $\beta_{k}$ and $\beta_{p}$, is able to enhance the geometric mechanism of production that we investigated. This necessarily implies that the squeezing of cosmological perturbations may affect perturbative particle production, and thus entanglement creation across the horizon. However, an important caveat to keep in mind is that a proper definition of particle states in an expanding background is not possible in general. 

For slowly expanding spacetimes, a solution to this issue is usually found by introducing the notion of \emph{adiabatic vacuum} \cite{gpp2}. Unfortunately, this technique is no longer reliable in case of a rapidly expanding quasi-de Sitter background. As discussed in Sec. \ref{sec2}, a reasonable in-vacuum state for inflationary fluctuations is represented by the Bunch-Davies vacuum, which represents a local attractor in the space of initial states for an expanding background. However, a proper definition of out states is usually problematic without assuming a transition to reheating period \cite{reh1,reh2,reh3} or to a radiation dominated phase at the end of inflation.  Accordingly, the notion of squeezed states during inflation is subject to ambiguities, as recently pointed out \cite{sq6}. We also remark that the here-depicted perturbative approach fails when the interaction term in Eq. \eqref{smat} becomes sufficiently large and thus comparable with the zero-order term of Dyson's expansion. 

As we can see from Figs. \ref{figquart}-\ref{figquad}, this is not the case of our model, since probability amplitudes of superhorizon pair production from perturbations are typically small in the slow-roll phase, provided some IR cutoff is properly applied. Such cutoff scale is ultimately related to fixing the initial value of the quasi-de Sitter scale factor, that we assumed as $a(\tau_i)=1/H_I$. A different choice, namely $a(\tau_i)=1$, is performed in Ref. \cite{brand20,msent4}, yielding a larger amount of entanglement entropy, despite leading to a high cutoff for super-Hubble modes. 

Finally, we expect that some back-reaction and/or decoherence mechanisms should take over at the end of inflation, thus modifying the overall scenario \cite{backr}. In Ref. \cite{bel1}, it was shown that classical back-reaction mechanisms are not expected to affect entanglement from inhomogeneities during slow-roll, but semiclassical effects and possible couplings of the inflaton to other quantum fields during reheating would presumably alter this picture. 

Incorporating these effects into the calculations of entropy for cosmological perturbations would undoubtedly aid in comprehending whether certain quantum ``signatures" of primordial perturbations managed to persist until the Cosmic Microwave Background radiation era.

%%%%%%%%%%%%%%%%%%%%%%%%%%%%%%%%%%%%%%%%%%%%%%%%%%%%%%%%%%%%%%%%%%%%%%%%%%%%%%%%

\section{Conclusion} \label{sec6}

In this work, we investigated entanglement between sub- and super-Hubble modes in the context of inflationary particle production from spacetime perturbations. We employed momentum-space entanglement techniques to show that a significant amount of entanglement entropy can be generated across the Hubble horizon by perturbative particle production during the slow-roll phase. Moreover, we noticed that such entropy is expected to grow rapidly for a quasi-de Sitter background evolution.

Inspired by the results of Planck satellite, we first considered a large-field inflationary scenario, invoking a quartic self-coupling potential and including a nonminimal coupling term with the scalar curvature of spacetime. Afterwards, we compared these outcomes with the well-known quadratic chaotic model of inflation and a small-field quadratic hilltop scenario, preserving field-curvature coupling in both cases. We showed that the amount of superhorizon entanglement from inhomogeneities is typically negligible in small-field scenarios, differently from large-field cases. 

Further, we discussed our findings in light of recent results for the squeezing entropy of cosmological perturbations, cubic nonlinearities and spectator scalar fields in inflation. In our analysis, we focused on the pure geometric contribution, neglecting the role of expansion, and thus the squeezing of perturbations, which in principle can enhance the total geometric production and, consequently, the entanglement entropy. 

Accordingly, the natural future step for our work would include this contribution into the probability amplitude for geometric production, despite the concept of inflationary squeezing is still under debate, due to the difficulties in defining particle states for a rapidly expanding background. Hence, generalizations of our study to the case of spectator fields are also possible. In particular, we expect that entanglement production for Dirac fields may lead to a different mode dependence of entanglement, due to the different statistics involved. Finally, the inclusion of back-reaction effects and decoherence mechanisms at the end of inflation may shed further light on the quantum properties of fluctuations and the corresponding quantum-to-classical transition of inflationary perturbations.

%%%%%%%%%%%%%%%%%%%%%%%%%%%%%%%%%%%%%%%%%%%%%%%%%%%%%%%%%%%%%%%%%%%%%%%%%%%%%%%%

\section*{Acknowledgements}
The work of OL is  partially financed by the Ministry of Education and Science of the Republic of Kazakhstan, Grant: IRN AP19680128. SM acknowledges financial support from ``PNRR MUR project PE0000023-NQSTI”.

\end{document}